\documentclass{article}
\usepackage[utf8]{inputenc}
\usepackage{amsmath}
\usepackage{amsfonts}
\usepackage{amssymb}
\usepackage{amsthm}

\def\p{\partial}

\newtheorem{prop}{Proposition}
\theoremstyle{remark}

\newtheorem{Rem}{Remark}
\newcommand{\dbar}{\bar{\partial}}
\newcommand{\wt}{\widetilde}
\newcommand{\be}{\begin{equation}}
\newcommand{\ee}{\end{equation}}
\newcommand{\bea}{\begin{eqnarray}}
\newcommand{\eea}{\end{eqnarray}}
\newcommand{\beaa}{\begin{eqnarray*}}
\newcommand{\eeaa}{\end{eqnarray*}}

\newcommand{\nn}{\nonumber}

\usepackage{authblk}
\author[1]{L.V. Bogdanov \thanks{leonid@itp.ac.ru}}
\author[2]{M.V. Pavlov}
\affil[1]{L.D. Landau ITP, Moscow}
\affil[2]{
Lebedev Physical Institute, Moscow
}

\title{Six-dimensional heavenly equation. Dressing scheme and the hierarchy.}
\date{}

\begin{document}

\maketitle

\begin{abstract}
We consider six-dimensional heavenly equation as
a reduction in the framework
of general six-dimensional linearly degenerate dispersionless hierarchy.
We characterise the reduction in terms 
of wave functions, introduce
generating relation,
Lax-Sato equations and develop the dressing scheme 
for the reduced hierarchy. Using the dressing scheme,
we construct a class of solutions for six-dimensional heavenly equation in terms of implicit functions.
\end{abstract}
\section{Introduction}
Six-dimensional heavenly equation  \cite{Takasaki89,PlebPrzan96}
\begin{equation}
\Theta_{x w} - \Theta_{y z} - \{ \Theta_x, \Theta_y\}_{(q,p)} = 0, 
\label{6DH}
\end{equation}
where 
$\theta = \theta (x, y, z, w, p, q)$,
$\{ \cdot , \cdot \}_{(q,p)}$ is the Poisson bracket          
$
\{ f_1, f_2\}_{(q,p)} := \frac{\partial f_1}{\partial q} 
\frac{\partial f_2}{\partial p}  -  \frac{\partial f_1}{\partial p} 
\frac{\partial f_2}{\partial q}, 
$
belongs to the class of quasiclassical self-dual Yang-Mills equations (SDYM equations for the Lie algebra of vector fields) and corresponds to the case of two-dimensional
Hamiltonian vector fields. Four-dimensional reductions of this equation include different versions of heavenly equation and related equations \cite{Takasaki89,PlebPrzan96,DF2010}.

It can be obtained from a standard SDYM type  Lax pair (taken in a special gauge)
\bea
&&
L=\p_{z}-\lambda \p_{x}+A_1,
\nn
\\
&&
M=\p_{w}-\lambda \p_{y}+A_2,
\label{YMpair}
\eea 
where $A_1$, $A_2$ (gauge field components) generally belong to some Lie algebra,
$\lambda$ is a complex variable (spectral parameter).
In our case $A_1$, $A_2$ are two-dimensional Hamiltonian vector fields.

Commutativity condition for operators (\ref{YMpair}) implies the
existence of potential $F$ (belonging to Lie algebra), $A_1=\p_{x}F$, $A_2=\p_{y}F$,
satisfying the equation
\bea
\p_{z}\p_{y}F-\p_{w} \p_{x}F-
[\p_{x}F,\p_{y}F]=0.
\label{YM}
\eea 
Six-dimensional heavenly equation case corresponds to two-dimensional Hamiltonian
vector fields $F$,
\beaa
F=\{\Theta,\dots\}_{(q,p)}:=\Theta_q\partial_p
-\Theta_p\partial_q,
\eeaa
and Lax pair (\ref{YMpair}) takes the form
\bea
&&
L=\p_{z}-\lambda \p_{x}+\{\Theta_x,\dots\}_{(q,p)},
\nn
\\
&&
M=\p_{w}-\lambda \p_{y}+\{\Theta_y,\dots\}_{(q,p)}.
\label{YMpairH}
\eea 
General properties of Lax pairs (more generally, involutive
distributions) of this type were discussed in \cite{BP17} (see also \cite{ManSan2006}, \cite{MS2014} and references therein, \cite{MarSer2012}).
According to Frobenius theorem 
for vector fields, linear equations
\bea
&&
L\Psi=(\p_{z}-\lambda \p_{x})\Psi+\{\Theta_x,\Psi\}_{(q,p)}=0,
\nn
\\
&&
M\Psi=(\p_{w}-\lambda \p_{y})\Psi+\{\Theta_y,\Psi\}_{(q,p)}=0
\label{YMpairHlin}
\eea
have four functionally independent solutions. Due to the special structure of vector fields, two of them are trivial,
\bea
\phi^1=x+\lambda z, \quad \phi^2=y+\lambda w,
\label{phi}
\eea
and two others can be found in generic form of series
in $\lambda$,
\bea
\Psi^1&=&q + 
\sum_{n=1}^\infty \Psi^1_n(p,q,x,y,z,w)\lambda^{-n},
\nn\\
\Psi^2&=&p + 
\sum_{n=1}^\infty \Psi^1_n(p,q,x,y,z,w)\lambda^{-n},
\label{Psi}
\eea
which for the case of Hamiltonian vector fields 
are canonically conjugate, 
$$
\{\Psi^1,\Psi^2\}_{(q,p)}=1.
$$

This is not a unique admissible form of series for 
solutions of linear equations (\ref{YMpairHlin}), they
can be also constructed as series in nonnegative powers
of $\lambda$.  Different basic sets of solutions of linear equations (\ref{YMpairHlin}) should be connected by
diffeomorphism, and this observation leads to formulation
of the dressing scheme based on Riemann-Hilbert problem
with relation between holomorphic components defined by
diffeomorphism (see below).

It is possible to introduce higher times and extend the
Lax pair (\ref{YMpairHn}) to the hierarchy of the form
\bea
&&
L_n=\p_{z_n}-\lambda^n \p_{x}+\{(H_{n-1})_x,\dots\}_{(q,p)},
\nn
\\
&&
M_n=\p_{w_n}-\lambda^n \p_{y}+\{(H_{n-1})_y,\dots\}_{(q,p)}.
\label{YMpairHn}
\eea 
where $H_{n-1}$ are polynomials in $\lambda$ of the order
$n-1$. The involutive distribution with the basis 
(\ref{YMpairH}), (\ref{YMpairHn}) is of codimension four
and the number of independent solutions of linear equations
of the form (\ref{YMpairHlin}) defined by this distribution remains equal to four; functions (\ref{Psi}) retain their form, and (\ref{phi}), (\ref{Psi}) are slightly 
modified to take into account the higher times,
\bea
\phi^1=x+\sum_{n=1}^{\infty}\lambda^n z_n, \quad \phi^2=y+\sum_{n=1}^{\infty}\lambda^n w_n.
\label{phiN}
\eea

The coefficients of the polynomials $H_{n-1}$
are connected with $\Theta$ by
commutativity conditions of the higher flow with initial
L,M operators. These conditions provide closed systems
of equations, the systems $[M_n,L]$, $[L_n,M]$ are
six-dimensional, and the systems $[L_n,L]$, $[M_n,M]$
are five-dimensional. This fact is connected with the degeneracy
and special structure 
of wave functions (\ref{phiN}) for 
linearly degenerate quasiclassical YM type hierarchies
\cite{BP17}, leading to simultaneous appearance of systems of
different dimensionalities in the same framework.
If we consider generic linearly degenerate 
six-dimensional dispesionless
hierarchy \cite{BDM},\cite{LVB09}, from which 
six-dimensional heavenly equation case 
can be obtained as a
reduction  (see below), this degeneracy
disappears.


A general feature of quasiclassical
self-dual YM type hierarchies \cite{BP17} is that
the basis of linear operators for the
higher flows of the hierarchy (\ref{YMpairHn}) can be represented
in compact recursive form,
\begin{align}
\wt L_n &:=L_{n+1}-\lambda L_n=\p_{z_{n+1}}-\lambda \p_{z_n}+\{\Theta_{z_n},\dots\}_{(q,p)},
\nn\\
\wt M_n &:=M_{n+1}-\lambda M_n=\p_{w_{n+1}}-\lambda \p_{w_n}+\{\Theta_{w_n},\dots\}_{(q,p)}.
\end{align}
It is interesting to note that each of the operators
$\wt L_k$, $\wt M_m$ is exactly of the same form
as $L$, $M$ operators (\ref{YMpairH}) (for another 
set of variables), and a commutator of arbitrary pair
of operators $\wt L_k$, $\wt M_m$ gives heavenly equation
(\ref{6DH}) for the respective set of variables,
\beaa
\Theta_{z_{n} w_{n+1}} - \Theta_{w_n z_{n+1}} - 
\{ \Theta_{z_n}, \Theta_{w_n}\}_{(q,p)} = 0
\eeaa
representing kind of intertwining equation for higher
flows of the hierarchy. This phenomenon 
is known for Yang-Mills type hierarchies both in the
standard \cite{AT93} and quasiclassical 
\cite{Takasaki89,BP17} case.
\section{The hierarchy. General six-dimensional case}
We will describe six-dimensional heavenly equation
hierarchy as a reduction of general
six-dimensional linearly degenerate
hierarchy and construct generating
equations and Lax-Sato equations for 6D heavenly equation hierarchy.

First we will briefly outline the picture 
of linearly degenerate
hierarchy developed in \cite{BDM}, \cite{LVB09} 
(see also \cite{BP17}). 
This picture starts from 
introducing the formal series for the wave functions,
defining a Pl\"ucker form which is a dual object to the
distribution of vector fields, and 
formulating a generating
equation for the hierarchy through the holomorphic properties of this form.

Let us consider the series 
\bea
\Psi^k=\Psi^k_0+\wt\Psi^k,
\quad
\Psi^k_0=\sum_{n=0}^\infty t^k_n \lambda^{n},\quad
\wt\Psi^k=\sum_{n=1}^\infty \Psi^k_n(\mathbf{t}^1,\dots,\mathbf{t}^{4})\lambda^{-n},
\label{levelform}
\eea
where $1\leqslant k\leqslant 4$ (for six-dimensional
hierarchy case), depending on four infinite sequences of independent variables
$\mathbf{t}^k=(t^k_0,\dots,t^k_n,\dots)$, $t^k_0=:x^k$.

The hierarchy is generated by the relation
\bea
\left( J^{-1}d\Psi ^{1}\wedge d\Psi ^{2}\wedge
d\Psi^{3}\wedge
d\Psi^{4}\right) _{-}=0,
\label{Gen}
\eea
where for linearly degenerate case the differentials do not
take into account $\lambda$ (considered as a parameter), and
vector fields of respective distribution do not contain
derivative over $\lambda$.
Here $\left( {\cdots}\right) _{-}$ denotes 
the projection to the part of 
$\left({\cdots}\right)$ 
with negative powers in $\lambda$ 
(respectively  $\left( {\cdots}\right) _{+}$ projects
to nonnegative powers)
and
$
J=\det
(\p_{x^j}\Psi^{i})_{i,j=1,\dots,4}.
$
Relation (\ref{Gen}) implies
Lax-Sato equations defining the dynamics of wave functions
over higher times, moreover, it is equivalent to the set of
Lax-Sato equations \cite{LVB09}.

Introducing the Jacobian matrix 
\begin{equation*}
(\text{Jac}_0)=\left(\frac{D(\Psi^1,\dots,\Psi^{4})} {D{({x^1,\dots,x^{4}})}}%
\right),\quad \det(\text{Jac}_0)=J,
\end{equation*}
it is possible to write the hierarchy in the Lax-Sato form, 
\begin{eqnarray}
&& \partial_{t^k_n}\mathbf{\Psi}=\sum_{i=1}^{4} \left((\text{Jac}_0)^{-1})_{ik}
\lambda^n)\right)_+ {\partial_i}\mathbf{\Psi},\quad 1\leqslant k \leqslant m,
\label{genSato}
\end{eqnarray}
where $1\leqslant n < \infty$, $\mathbf{\Psi}=(\Psi^1,\dots,\Psi^{4})$, $\p_i=\p_{x^i}$.


First flows of the hierarchy (lowest level integrable distribution) read 
\begin{eqnarray}
\partial_{t^k_1}\mathbf{\Psi}=(\lambda \partial_k-\sum_{p=1}^{4} (\partial_k
u_p)\partial_p)\mathbf{\Psi},\quad 1\leqslant k\leqslant 4,
\label{genlinear}
\end{eqnarray}
where $u_k=\Psi^k_1$. 

To proceed to the case of six-dimensional heavenly equation hierarchy, we will need a
reduction $J=1$ corresponding to volume-preserving (divergence-free
vector fields) case, generating relation in this case is
\bea
\left(d\Psi ^{1}\wedge d\Psi ^{2}\wedge
d\Psi^{3}\wedge
d\Psi^{4}\right) _{-}=0,
\label{GenV}
\eea
vector fields (\ref{genlinear}) are divergence-free,
$\sum \p_p u_p=0$.

\section{The hierarchy. Description of reduction}
To obtain six-dimensional heavenly equation hierarchy,
we consider a reduction of
the hierarchy (\ref{GenV})
characterised by the condition that two
of the series $\Psi^k$
are equal to `vacuum' functions $\Psi^k_0=\sum_{n=0}^\infty t^k_n\lambda^n$ (for finite subsets of times they are
polynomial)
\be 
(\Psi^3)_-=0,\; \Psi^3=\Psi^3_0=\sum_{n=0}^\infty t^k_n\lambda^n,
\quad 
(\Psi^4)_-=0,\; \Psi^3=\Psi^3_0=\sum_{n=0}^\infty t^k_n\lambda^n,
\label{reduction}
\ee 
Generating relation for six-dimensional 
heavenly equation hierarchy case is
\beaa
\left(d\Psi ^{1}\wedge d\Psi ^{2}\wedge 
d\Psi^{3}_0
\wedge d\Psi
^{4}_0 \right) _{-}=0,
\eeaa
We will restrict ourselves to the higher flows of
six-dimensional 
heavenly equation type and drop higher times in $\Psi^1$, $\Psi^2$,
also introducing new notations for functions $\Psi^3$, $\Psi^4$ and corresponding times
to make the calculations more transparent.
Thus we consider the wave functions of the form
\bea
&&
\Psi^1=q+\sum_{n=1}^{\infty}\Psi^1_n \lambda^{-n},
\quad
\Psi^2=p+\sum_{n=1}^{\infty}\Psi^2_n \lambda^{-n},
\\&&
\phi^1:=\Psi_0^{3}=\sum_{n=0}^\infty z_n \lambda^n,
\quad x:=z_0,\; z:=z_1
\\&&
\phi^2:=\Psi_0^{4}=\sum_{n=0}^\infty w_n \lambda^n,
\quad y:=w_0,\; w:=w_1
\label{psiphi}
\eea
where $q$, $p$, $z_n$, $w_n$ are independent variables 
and coefficients $\Psi^1_n$, $\Psi^2_n$ are considered as
dependent variables, 
generating relation for six-dimensional 
heavenly equation hierarchy reads
\bea
\left(d\Psi ^{1}\wedge d\Psi ^{2}\wedge 
d\phi^{1}
\wedge d\phi
^{2} \right) _{-}=0.
\label{Gen6}
\eea
Lax-Sato equations (\ref{genSato}) have rather special structure
in this case, taking into account that the only nonzero entries of last two lines of the Jacobian
form the unity matrix:
\begin{gather}
\p_{z_n}\mathbf\Psi=
\left(
\lambda^n\p_{x} + 
(\lambda^n\{\Psi^1,\Psi^2\}_{(p,x)})_+\p_q
-
(\lambda^n\{\Psi^1,\Psi^2\}_{(q,x)})_+\p_p
\right)
\mathbf\Psi,
\nn\\
\p_{w_n}\mathbf\Psi=
\left(
\lambda^n\p_{y} + 
(\lambda^n\{\Psi^1,\Psi^2\}_{(p,y)})_+\p_q
-
(\lambda^n\{\Psi^1,\Psi^2\}_{(q,y)})_+\p_p
\right)
\mathbf\Psi,
\nn\\
\{\Psi^1,\Psi^2\}_{(q,p)}=1,\quad
\label{LS6}
\end{gather}
here $\mathbf\Psi= (\Psi^1,\Psi^2)$, 
$
\{f,g\}_{(x,y)}:=f_x g_y-f_y g_x
$
Vector fields are Hamiltonian due to the 
condition
$
\{\Psi^1,\Psi^2\}_{(q,p)}=1
$.
Lax-Sato equations define the evolution of the series $\Psi^1$,
$\Psi^2$ with the coefficients considered as functions of four variables 
$(q,p,x,y)$ with respect to the higher times. The first
two flows of the hierarchy read
\bea
&&
\p_{z}\mathbf\Psi=
\left(
\lambda\p_{x} 
-(\p_x\Psi^1_1)\p_q
-
(\p_x\Psi^2_1)\p_p
\right)
\mathbf\Psi,
\nn\\
&&
\p_{w}\mathbf\Psi=
\left(
\lambda\p_{y}  
-(\p_y\Psi^1_1)\p_q
-
(\p_y\Psi^2_1)\p_p
\right)
\mathbf\Psi,
\label{LS60}
\eea
and the first nontrivial order of expansion in $\lambda$
of the condition 
$
\{\Psi^1,\Psi^2\}_{(q,p)}=1
$
gives
$
\p_q \Psi^1_1 + \p_p \Psi^2_1=0,
$
thus implying the existence of the potential $\Theta$,
$\Psi^1_1=-\Theta_p$, $\Psi^2_1=\Theta_q$, transforming
equations (\ref{LS60})  to the form corresponding to
the Lax pair of six-dimensional 
heavenly  equation (\ref{YMpairH}),
\beaa
&&
(\p_{z}-
\lambda \p_{x})\mathbf{\Psi}
=\{\mathbf{\Psi},\Theta_x\}_{(q,p)},
\nn
\\
&&
(\p_{w}
-\lambda \p_{y})\mathbf{\Psi}
=
\{\mathbf{\Psi},\Theta_y\}_{(q,p)}.
\eeaa  
Higher flows (\ref{LS6}) can be written in a simple recursive form,
which can be also obtained directly from the generating
relation (\ref{Gen6}) (see \cite{BP17} for more detail)
\begin{align}
(\p_{z_{n+1}}-\lambda \p_{z_n})\mathbf{\Psi}
&=\{\mathbf{\Psi},\Theta_{z_n}\}_{(q,p)},
\nn\\
(\p_{w_{n+1}}-\lambda \p_{w_n})\mathbf{\Psi}
&=\{\mathbf{\Psi},\Theta_{w_n}\}_{(q,p)}.
\label{recur}
\end{align}
A comparison between Lax-Sato equations (\ref{LS6})
and recursive relations (\ref{recur}) provides useful expressions of coefficients of expansion of Poisson
brackets in $\lambda$ through the derivatives of potential
$\Theta$.
\begin{Rem}
6D heavenly equation and the hierarchy can be easily
generalized to the case of multidimensional Poisson bracket and Hamiltonian vector fields, in which the basic equation (connected
to hyper-K\"ahler equations \cite{Takasaki89}) reads
\begin{equation}
\Theta_{x w} - \Theta_{y z} - \{ \Theta_x, \Theta_y\}_{(\mathbf{q},\mathbf{p})} = 0, 
\label{YMHE}
\end{equation}
where 
$$\{f,g\}{(\mathbf{q},\mathbf{p})}:=\sum_{i=1}^{N}
\frac{\partial f_1}{\partial q_i} 
\frac{\partial f_2}{\partial p_i}  -  \frac{\partial f_1}{\partial p_i} 
\frac{\partial f_2}{\partial q_i},
$$
and the generating relation for the hierarchy is
\beaa
\left(\left(
\sum_{k=0}^{N-1} d\Psi ^{2k+1}\wedge d\Psi ^{2k+2}
\right)
\wedge 
d\phi^{1}
\wedge d\phi
^{2} \right) _{-}=0,
\eeaa
the generalization of Lax-Sato equations is straightforward.
\end{Rem}
\begin{Rem}
We have already mentioned (see also \cite{BP17}) 
that Yang-Mills type linearly degenerate dispersionless
hierarchies are rather special, and, if we take into account different sets of times (submanifolds in the space
of independent variables)
in the generating relation, may contain equations of different dimensionalities. Let us consider a generating
relation
\bea
\left((d\Psi ^{1}\wedge d\Psi ^{2})
\wedge (d\phi^{1}
\wedge d\phi
^{2}\wedge 
\dots\wedge d\phi ^{N})\right) _{-}=0,
\label{Gen6X}
\eea
where $N$ is arbitrary (may be infinite) and 
$\Psi^1$, $\Psi^1$ are of general form (\ref{levelform}), taking into higher times.
Evidently, this relation contains several copies of 6DHE
hierarchy considered in this work. It also contains
copies of standard 4-dimensional heavenly 
equation hierarchy. Morover, any solution of generating
relation (\ref{Gen6X}) for $N=P-1$ evidently gives
a solution for $N=P$.
Involutivity conditions
for the distribution corresponding to relation
(\ref{Gen6X}) after restriction to
some submanifolds of independent variables
lead to equations of higher dimensionalities
up to $N+4$ with the Lax pairs of the type
\beaa
&&
L=\p_{z}-\lambda \p_{x}+\{\Theta_x,\dots\}_{(q,p)},
\nn
\\
&&
\wt M=\p_{\wt w_N}+\sum_{n=1}^{N-1}\lambda^n \p_{\wt w_n}+
\{\wt H_N(\lambda),\dots\}_{(q,p)}.
\eeaa
where $H_N(\lambda)$ is a polynomial of the order
$N-2$ and the set of independent variables consists
of $z$, $x$, $p$, $q$, $\wt w_n$, 
$1\leqslant n \leqslant N$.
We believe that the hierarchy (\ref{Gen6X}) containing
subhierarchies of different dimensionalities may have
an interesting geometric interpretation, probably in
the language of exotic cohomologies developed in 
\cite{Morozov}.
\end{Rem}
\section{Dressing scheme and solutions}
Starting from the dressing scheme for general 
six-dimensional hierarchy (\ref{Gen}) 
\cite{BK2005}, \cite{BDM}, 
\cite{LVB09},  by the reduction to
the 6D heavenly equation 
hierarchy (\ref{Gen6}) (see also \cite{BP17})
we obtain Riemann-Hilbert
problem on the unit circle (or the boundary of some
region $G$)
\beaa
\Psi ^{1}_\text{in} =
F^1(\lambda,\Psi ^{1},\Psi^2;
\phi^{1},\phi^{2})_\text{out},
\\
\Psi ^{2}_\text{in} =
F^2(\lambda,\Psi ^{1},\Psi^2;
\phi^{1},\phi^{2})_\text{out},
\eeaa
where the diffeomorphism defined by $F_1$, $F_2$
for the case of Hamiltonian reduction should be
area-preserving with respect to the variables
$\Psi^1$, $\Psi^2$,
or the $\dbar$ problem in the unit disk (or
some region $G$)
\bea
&&
\dbar\Psi ^{1} =
W_{,2}(\lambda,\bar\lambda,\Psi ^{1},\Psi^2;
\phi^{1},\phi^{2}),\quad W_{,2}:=\frac{\p W}{\p \Psi^2},
\nn
\\
&&
\dbar \Psi ^{2} =
-W_{,1}(\lambda,\bar\lambda, \Psi ^{1},\Psi^2;
\phi^{1},\phi^{2}),\quad
W_{,1}:=\frac{\p W}{\p \Psi^1}.
\label{dbar}
\eea
Here the Hamiltonian reduction is taken into account 
explicitly.

The functional freedom of
the dressing data consists of functions of 5 variables,
that indicates that reduced equations are generically
6-dimensional, like equations of the unreduced 
linearly-degenerate dispersionless hierarchy.

We search for the solutions of the form
\beaa
\Psi^1=q+ \tilde \Psi^1,\quad \Psi^2=p+ \tilde \Psi^2
\eeaa
where $\tilde \Psi^1$, $\tilde \Psi^2$ are analytic outside $G$
and go to zero at infinity.

The $\dbar$ problem can be obtained by the variation of the action
\bea
f
=\frac{1}{2\pi\mathrm{i}}\iint_{G}
\left(\wt \Psi^2 \dbar \wt \Psi^1 -
W(\lambda,\bar \lambda,\Psi^1,\Psi^2;\phi^{1},\phi^{2})\right)d\lambda\wedge d\bar \lambda,
\label{action}
\eea
where one should consider independent variations of $\wt\Psi^1$, $\wt\Psi^2$
possessing required analytic properties, 
keeping times $q,p,\dots$ fixed. Using the results of the work 
\cite{BK2005} in our setting, we come to the following statement:
\begin{prop} 
The function
\bea
\Theta(q,p,\mathbf{z},\mathbf{w})=
\frac{1}{2\pi\mathrm{i}}\iint_{G}
\Bigl(\wt \Psi^2 \dbar \wt \Psi^1
-
W(\lambda,\bar \lambda,\Psi^1,\Psi^2;\phi^{1},\phi^{2})
\Bigr) 
d\lambda\wedge d\bar \lambda,
\label{HEtau}
\eea
i.e., the action (\ref{action}) evaluated on the solution
of the $\dbar$ problem (\ref{dbar}), gives the potential
for 6D heavenly equation hierarchy  
and satisfies 6D heavenly equation
(\ref{6DH}).
\end{prop}
To prove this proposition, it is enough to 
check the relations
$\Psi^1_1=-\partial_p\Theta$,
$\Psi^2_1=\partial_q \Theta$.
\subsection*{A class of solutions}
Below we will consruct a class of solutions for the 6DHE
hierarchy. We will
go along the lines of similar calculation for
general heavenly equation 
presented in \cite{LVB15}.

A class of solutions for the 6D heavenly equation hierarchy
(\ref{6DH})
in terms of implicit functions (similar to
to solutions of hyper-K\"ahler hierarchy
presented in
\cite{Gindikin86},
\cite{Takasaki89})
can be constructed using the choice
\beaa
&&
W(\lambda,\bar \lambda,\Psi^1,\Psi^2;\phi^1,\phi^2)=
\\
&&\quad
=2\pi\mathrm{i}\left(\sum_{i=1}^{M}\delta(\lambda-\mu_i)
F_i(\Psi^1;\phi^1,\phi^2)
+
\sum_{i=1}^{M}\delta(\lambda-\nu_i)
G_i(\Psi^2;\phi^1,\phi^2)\right),
\eeaa
where $\delta(\lambda-\mu_i)$, $\delta(\lambda-\nu_i)$ are two-dimensional
delta functions in the complex plane, and $F_i$, $G_i$ are some (complex-analytic) functions of
three variables. The $\dbar$ problem (\ref{dbar}) in this case reads
\bea
\dbar \Psi^1&=&2\pi\mathrm{i}\sum_{i=1}^{M}
\delta(\lambda-\nu_i)G'_i(\Psi^2;\phi^1,\phi^2),
\nn\\
\dbar \tilde \Psi^2&=&-2\pi\mathrm{i}\sum_{i=1}^{M}
\delta(\lambda-\mu_i)F'_i(\Psi^1;\phi^1,\phi^2).
\label{dbardelta}
\eea
The solutions of the $\dbar$ problem are then of
the form
\bea
\Psi^1=q+\sum_{i=1}^{M} \frac{f_i}{\lambda-\nu_i}, \quad  
\Psi^2=p+\sum_{i=1}^{M} \frac{g_i}{\lambda-\mu_i},
\label{psi}
\eea
and from (\ref{dbar}) the functions $f_i$, $g_i$ are defined as implicit functions,
\bea
f_i(q,p,\mathbf{z},\mathbf{w})&=&G'_i\left(p+
\sum_{k=1}^{M} \frac{g_k(q,p,\mathbf{z},\mathbf{w})}{\nu_i-\mu_k};
\sum_{n=0}^\infty z_n \nu_i^n,\sum_{n=0}^\infty 
w_n \nu_i^n
\right),
\nn\\
g_i(q,p,\mathbf{z},\mathbf{w})&=&-F'_i\left(q+
\sum _{k=1}^{M}\frac{f_k(q,p,\mathbf{z},\mathbf{w})}{\mu_i-\nu_k};
\sum_{n=0}^\infty z_n \mu_i^n,\sum_{n=0}^\infty 
w_n \mu_i^n\right).
\label{impl}
\eea 
The potential $\Theta$ solving the general heavenly equation hierarchy
is then given by the formula (\ref{HEtau}), it depends on the set
of arbitrary functions of three variables $F_i$, $G_i$,
\begin{multline} 
\Theta(\mathbf{x})=
\sum_{i=1}^{M} 
F_i(\Psi^1(\mu_i);\phi^1(\mu_i),\phi^2(\mu_i))
+
\sum_{i=1}^{M} 
G_i(\Psi^2(\nu_i);\phi^1(\nu_i),\phi^2(\nu_i))+
\\
+\sum_{i=1}^{M}\sum_{j=1}^{M} \frac{f_i g_j}{\nu_i - \mu_j},
\label{Theta}
\end{multline} 
where $\Psi^1$, $\Psi^2$ are given by
(\ref{psi}), $\phi^1$, $\phi^2$ have the form
(\ref{psiphi}),
and the functions $f_i$, $g_i$ are defined as implicit functions 
by equations (\ref{impl}). Formula (\ref{Theta}) corresponds
to the special solution of hyper-K\"ahler hierarchies 
presented
in \cite{Gindikin86}, \cite{Takasaki89}, however, it is
important to note that in our case the solution depends
on the set of arbitrary functions of three variables,
in contrast to the set of functions of one variable
in \cite{Gindikin86}, \cite{Takasaki89}.
\section*{Acknowledgements}
MVP's work was partially supported by the grant
of Presidium of RAS "Fundamental Problems
of Nonlinear Dynamics" and by the RFBR grant 17-01-00366.

\end{document}